\documentclass{article}
\usepackage{amssymb}

\usepackage{graphicx}
\usepackage{amsmath}

\newtheorem{theorem}{Theorem}

\newtheorem{corollary}[theorem]{Corollary}

\newtheorem{proposition}[theorem]{Proposition}

\newenvironment{proof}[1][Proof]{\textbf{#1.} }{\ \rule{0.5em}{0.5em}}
\input{tcilatex}

\begin{document}

\title{On Eigenvalues Problem for Self-adjoint Operators with Singular Perturbations}
\author{Sylwia Kondej \\
Institute of Theoretical Physics, \\
University of Wroclaw, Pl. Maxa Borna 9, Poland\\
e-mail: kondej@ift.uni.wroc.pl}
\maketitle

\begin{abstract}
We investigate the eigengenvalues problem for self-adjoint operators with
the singular perturbations. The general results presented here include
weakly as well as strongly singular cases. We illustrate these results on
two models which correspond to so-called additive strongly singular
perturbations.
\end{abstract}

\section{INTRODUCTION}

Ever since Kr\"{o}nig, Penney \cite{KrPe} and Bethe, Peierls \cite{BePe}
used potential supported by isolated points the Hamiltonians with
perturbations on null sets attracted physicists and mathematicians. Special
progress in understanding of mathematical aspects started in 80's. Most of
the researches has been devoted to study of the potentials supported by null
sets but it turns out that the perturbations by the dynamics of the systems
living on null sets are also interested (see for example \cite{AHH, AK}). In
this paper we are concerned with method of perturbations applicable in both
cases.

Let \emph{A} represents Hamiltonian of free system \emph{S} and $\mathbb{V}$
corresponds to a potential supported by a null set or to a Hamiltonian of
system located on a null set. There are many papers devoted to problem of
the construction of Hamiltonian of composite system \emph{S}. This problem
can be formulated as follows. Give a meaning of self-adjoint operator to the
following formal sum 
\begin{equation}
A+\mathbb{V}.  \label{A+V}
\end{equation}

In this paper we investigate the eigenvalues problem for the self-adjoint
operators with the singular perturbations.

Let us describe the idea of singular perturbations more precisely. The main
concept is based on the theory of the extensions of symmetric operators
developed by von Neuman and Krein.

Let \emph{A} be a self-adjoint, strictly positive operator in the Hilbert
space $\mathcal{H}$ and $D(A)$ denote its domain. We say that a
self-adjoint, invertible operator $\tilde{A}$ in $\mathcal{H}$ is a \emph{%
singular perturbation} of $A$\ if $\tilde{A}$ coincides with $A$ on a set
dense in $\mathcal{H}$. The class of all singular perturbation of \emph{A}
we denote by $\mathcal{A}_{s}(A).$ In other words, any operator $\tilde{A}%
\in \mathcal{A}_{s}(A)$\ is the self-adjoint extension of closed, symmetric
operator $\dot{A}\equiv A\lfloor D$ where $D\equiv D(A)\cap D(\tilde{A})$
and $\lfloor D$\ stands for the restriction to \emph{D.}\ 

In accordance with the standard notations, we put $N_{0}$\ for the
deficiency space of $\dot{A}$ which coincides with $\ker \dot{A}^{\ast }$
where $\dot{A}^{\ast }$ stands for the adjoint operator to $\dot{A}.$ It is
known \cite{AKK,KKO} that any self-adjoint extension $\tilde{A}$ of $\dot{A}%
, $ can be represented by its inverse in the following way 
\begin{equation}
\tilde{A}^{-1}=A^{-1}+\tilde{B}^{-1}  \label{geAtyl}
\end{equation}
where $\tilde{B}^{-1}:\mathcal{H}\rightarrow N_{0}$ is self-adjoint. Note
that (\ref{geAtyl}) is just the Krein formula at point zero.\ 

The class $\mathcal{A}_{s}(A)$\ can be decomposed into so-called weakly and
strongly singular classes \emph{i.e.} $\mathcal{A}_{s}(A)=\mathcal{A}%
_{ws}(A)\cup \mathcal{A}_{ss}(A).$\ The precise definitions of $\mathcal{A}%
_{ws}(A)$\ and $\mathcal{A}_{ss}(A)$\ we will be formulated later.

Some of operators from $\mathcal{A}_{s}(A)$ can be interpreted as the
self-adjoint realizations of (\ref{A+V}). But we would like to emphasize
that the sum (\ref{A+V})\ only in some remote sense corresponds to addition
and operator $\mathbb{V}$ does not act in $\mathcal{H}$. In fact, we usually
write 
\begin{mathletters}
\begin{equation}
\tilde{A}=A\tilde{+}\mathbb{V}\text{ \ \ \ \ \ for }\tilde{A}\in \mathcal{A}%
_{ws}(A)  \label{szok}
\end{equation}
and 
\end{mathletters}
\begin{equation}
\tilde{A}=A\hat{+}\mathbb{V}\text{ \ \ \ \ \ for }\tilde{A}\in \mathcal{A}%
_{ss}(A).  \label{sumdach}
\end{equation}
The constructions of (\ref{szok}) and (\ref{sumdach})\ were described in 
\cite{AKK2,TKVK,VK1,VK2} and \cite{SK1} respectively.

The aim of this paper is to describe some of the spectral properties of
operators from $\mathcal{A}_{s}(A).$\ Precisely, we are interested in the
problem of eigenvalues 
\begin{equation}
\tilde{A}f=Ef\text{ \ \ \ \ }E\in \mathbb{R}\backslash \{0\},\text{\ }f\in D(%
\tilde{A}).  \label{eiequ1}
\end{equation}

Our first goal is to characterize the solutions of (\ref{eiequ1})\ in the
terms of $\tilde{B}^{-1}=\tilde{A}^{-1}-A^{-1}.$ The eigenvalues problem for
operator given by $A\tilde{+}\alpha \mathbb{V\in }\mathcal{A}_{ws}(A)$\
where $\alpha \in \mathbb{R}$ was studied in \cite{AKK2}.\ Here we start
from the general situation and in further discussion focus on some subfamily
of $\mathcal{A}_{s}(A)$\ which we denote by $\mathcal{A}_{s}^{\prime }(A).$
However $\mathcal{A}_{s}^{\prime }(A)\cap \mathcal{A}_{ws}(A)\neq \{0\}$ as
well as $\mathcal{A}_{s}^{\prime }(A)\cap \mathcal{A}_{ss}(A)\neq \{0\}$
holds.\ 

In the second part of this paper we discuss (\ref{eiequ1}) for $\tilde{A}$
given by $A\hat{+}\alpha \mathbb{V\in }\mathcal{A}_{ss}(A).$ We consider two
models in which \emph{A}\ corresponds to the Laplace operators in $%
L^{2}(\left\langle 0,\pi \right\rangle \times \mathbb{R}^{2},dx)$ and $L^{2}(%
\mathbb{R}^{3},dx)$ and $\mathbb{V}$\ refers to self-adjoint operators in $%
L^{2}(I,dx_{1})$ and $L^{2}(C,d\phi )$ where \emph{I, C} denote an interval
and circle respectively. In particular we investigate the asymptotic
behaviour of the solutions of (\ref{eiequ1}) as $\alpha \rightarrow \infty .$

\subsection{Definitions and Notations}

Let $A$ be a self-adjoint, strictly positive operator in separable Hilbert
space $\mathcal{H}$ with an inner product $(\cdot ,\cdot )$ and norm $%
\left\| \cdot \right\| $. By $D(A)$ and $Ran(A)$ we denote the domain and
range of $A$.

Put $\rho (A)$ and $\sigma (A)$ for the resolvent set and spectrum of $A$
respectively. In this paper we assume for simplicity that the spectrum of $A$
is purely absolutely continuous, $\sigma (A)=\sigma _{ac}(A).$

Let $q$ be integer with $\mid q\mid \leq 2$. Define the inner product in $%
D(A)$ by 
\begin{equation*}
(u,v)_{q}=(A^{q/2}u,A^{q/2}v).
\end{equation*}

Putting $\mathcal{H}_{q}$ for the completion of $D(A)$ in the norm $\left\|
\cdot \right\| _{q}$ we get the chain of the Hilbert spaces 
\begin{equation}
\mathcal{H}_{-2}\supset \mathcal{H}_{-1}\supset \mathcal{H}_{0}\equiv 
\mathcal{H}\supset \mathcal{H}_{1}\supset \mathcal{H}_{2}.  \label{cha}
\end{equation}
Clearly, for \emph{q=1, 2} the space $\mathcal{H}_{q}$ coincides with $%
D(A^{q/2}).$

From the construction of (\ref{cha}) follows that $\mathcal{H}_{q}$ and $%
\mathcal{H}_{-q}$ are mutually conjugate with respect to $\mathcal{H}_{0}.$
Set $\langle \cdot ,\cdot \rangle $ for the duality between $\mathcal{H}_{q}$
and $\mathcal{H}_{-q}$.

In chain (\ref{cha}) the operator $A$ is unitary as a map from $\mathcal{H}%
_{2}$ to $\mathcal{H}_{0}$ and it acts isometrically from $\mathcal{H}_{1}$
to $\mathcal{H}_{-1\text{ }}$and from $\mathcal{H}_{0}$ to $\mathcal{H}_{-2}$%
. Putting $\mathbb{A}$ for the closure of $A:\mathcal{H}_{0}\rightarrow 
\mathcal{H}_{-2}$ we obtain unitary operator.

By definition, a self-adjoint, invertible operator $\widetilde{A}$ in $%
\mathcal{H}$ is called a singular perturbation of $A$ if the set 
\begin{equation}
\mathcal{D}=\{\varphi \in D(A)\cap D(\widetilde{A}):\text{ }A\varphi =%
\widetilde{A}\varphi \}  \label{set}
\end{equation}
is dense in $\mathcal{H}$. The set of all singular perturbations of \emph{A}
will be denoted by $\widetilde{A}\in \mathcal{A}_{s}(A).$

So, we see that any $\widetilde{A}\in \mathcal{A}_{s}(A)$ is a self-adjoint
extension of symmetric operator 
\begin{equation*}
\overset{\cdot }{A}=A\lfloor \mathcal{D}
\end{equation*}
which is automatically closed. Then $\mathcal{D}$ is the proper subspace of $%
\mathcal{H}_{2}$. Denote $M_{2}=\mathcal{D}$ and $N_{2}=\mathcal{H}%
_{2}\circleddash M_{2}.$ Conversely, if a linear space $M_{2}\subset 
\mathcal{H}_{2}$ is closed in $\mathcal{H}_{2}$\ and dense in $\mathcal{H}%
_{0}$ then $\overset{\cdot }{A}=A\lfloor M_{2}$\ is symmetric and its
self-adjoint extension belongs to $\mathcal{A}_{s}(A).$\ 

By unitarity of $A:\mathcal{H}_{2}\rightarrow \mathcal{H}_{0}$ we can
''shift'' the above decomposition of $\mathcal{H}_{2}$ onto $\mathcal{H}_{0}$
$i.e.$%
\begin{equation*}
\mathcal{H}_{0}=M_{0}\oplus N_{0}
\end{equation*}
where $M_{0}=AM_{2}$, $N_{0}=AN_{2}.$ Note that $N_{0}$ coincides with $\ker 
\dot{A}^{\ast }$ and is the defect space for $\overset{\cdot }{A}$.

We shall say that operator $\tilde{B}^{-1}$ belongs to $\mathcal{\tilde{B}}%
(A)$ if there exists a proper subspace $N$ of $\mathcal{H}_{0}$ so that $%
M_{+}\equiv \mathcal{H}_{2}\ominus A^{-1}N$ is dense in $\mathcal{H}_{0}$
and 
\begin{equation*}
\tilde{B}^{-1}=0\oplus B^{-1}:M\oplus N\rightarrow \mathcal{H}_{0},\text{ \
\ \ }M\equiv AM_{+}
\end{equation*}
where $B^{-1}$ is self-adjoint, invertible in $N.$ The following theorem
states the relation between $\mathcal{\tilde{B}}(A)$ and $\mathcal{A}_{s}(A)$%
.

\begin{theorem}[\protect\cite{AKK,KKO}]
If $\tilde{B}^{-1}\in \mathcal{\tilde{B}}(A)$ then 
\begin{equation}
\widetilde{A}\text{ }^{-1}=A^{-1}+\widetilde{B}\text{ }^{-1}  \label{kko}
\end{equation}
is invertible and $\widetilde{A}\in \mathcal{A}_{s}(A)$.\ Moreover $%
\widetilde{A}\lfloor \ker A^{-1}\tilde{B}^{-1}=\dot{A}.$ Conversely, if $%
\widetilde{A}\in \mathcal{A}_{s}(A)$ then there exists exactly one $\tilde{B}%
^{-1}\in \mathcal{\tilde{B}}(A)$ so that (\ref{kko}) holds and $kerA^{-1}%
\tilde{B}^{-1}=D(\dot{A})$.
\end{theorem}

In $\mathcal{A}_{s}(A)$ we can select weakly singular class defined by 
\begin{equation*}
\mathcal{A}_{ws}(A):=\{\tilde{A}\in \mathcal{A}_{s}(A):D(\tilde{A})\subset 
\mathcal{H}_{1}\}
\end{equation*}
and strongly singular class 
\begin{equation*}
\mathcal{A}_{ss}(A):=\mathcal{A}_{s}(A)\backslash \mathcal{A}_{ws}(A).
\end{equation*}

\section{Eigenvalues problem for singular perturbation}

We keep notations introduced in the previous section. Now, our goal will be
to investigate eigenvalues problem for $\widetilde{A}\in \mathcal{A}_{s}(A)$%
. Precisely, we would like to formulate conditions ensuring the existence of
solutions 
\begin{equation}
\widetilde{A}f=Ef\text{ },\text{ \ \ }f\in D(\widetilde{A}),\text{ }E\in 
\mathbb{R}\backslash \{0\}.  \label{wy0}
\end{equation}
The pure point spectrum of $\tilde{A}$ \emph{i.e.} the set of all \emph{E}
satisfying (\ref{wy0}) we denote by $\sigma _{p}(\tilde{A}).$

Due to theorem 1 for any $\widetilde{A}\in \mathcal{A}_{s}(A)$ there exists
uniquely defined $\tilde{B}^{-1}\in \mathcal{\tilde{B}}(A)$ so that $%
\widetilde{A}$ $^{-1}=A^{-1}+\tilde{B}^{-1}$. Thus (\ref{wy0}) can be
rewritten in the form 
\begin{equation}
(A^{-1}+\tilde{B}^{-1})g=E^{-1}g,\text{ \ \ \ \ }g=\widetilde{A}f,\text{ }%
E\in \mathbb{R}\backslash \{0\}.  \label{wyp}
\end{equation}

For all $E\in \mathbb{R}\backslash \{0\}$ we define $U_{0E}=(A-E)A^{-1}:%
\mathcal{H}\rightarrow \mathcal{H}$, $\mathcal{H}\equiv \mathcal{H}_{0}$.
With this notation we can formulate the following theorem.

\begin{theorem}
A pair $E\in \mathbb{R}\backslash \{0\},$\ $f\in \mathcal{H}$ solves eq.(\ref
{wy0}) iff f has the form 
\begin{equation*}
f=\tilde{A}^{-1}g
\end{equation*}
where g$\in \ker (U_{0E}-E\tilde{B}^{-1}).$
\end{theorem}

\begin{proof}
Note that (\ref{wyp}) is equivalent to 
\begin{equation}
(\mathbb{I}-EA^{-1})g-E\tilde{B}^{-1}g=0,\text{ \ }g=\tilde{A}f.
\label{pom1}
\end{equation}
In turn, (\ref{pom1}) can be written as 
\begin{equation*}
U_{0E}g-E\tilde{B}^{-1}g=0,\text{ \ }g=\tilde{A}f.
\end{equation*}
So the theorem is proven.
\end{proof}

It is not hard to see that theorem 2 is generalization of theorem 5 \cite
{AKK2} but the proof presented above is less complicated.

To discuss eigenvalues problem (\ref{wy0}) in more details we shall consider
cases $E\in \rho (A)\cap \mathbb{R}$ and $E\in \sigma (A)$ separately.

First, assume $E\in \rho (A)\cap \mathbb{R}$. Define operator-valued
function on $\rho (A)\cap \mathbb{R}$ by $R_{E}:=(A-E)^{-1}:\mathcal{H}%
\rightarrow \mathcal{H}.$ Then $U_{E0}\equiv (U_{0E})^{-1}=AR_{E}:\mathcal{H}%
\rightarrow \mathcal{H}$. Let $\mathcal{K}_{E}$ stand for the set of
eigenvectors of $U_{E0}\tilde{B}^{-1}$.

Given $E\in \rho (A)\cap \mathbb{R}$ and $h\in \mathcal{K}_{E}$ the
expression 
\begin{equation*}
a_{h}(E^{\prime })\equiv a_{h,E}(E^{\prime }):=\left\| h\right\|
^{-2}(U_{E^{\prime }0}\tilde{B}^{-1}h,h)
\end{equation*}
defines a real, continous function on $\rho (A)\cap \mathbb{R}$.

Equivalently 
\begin{eqnarray*}
a_{h}(E^{\prime }) &=&\left\| h\right\| ^{-2}\left\langle (A-E^{\prime
})^{-1}\tilde{B}^{-1}h,\mathbb{A}h\right\rangle = \\
&&\left\| h\right\| ^{-2}\int_{\lambda }^{\infty }\frac{1}{t-E^{\prime }}%
d\left\langle \mathbb{E}_{t}\tilde{B}^{-1}h,\mathbb{A}h\right\rangle
\end{eqnarray*}

where $\lambda >0$ is the lower bound for \emph{A} and $\mathbb{E}_{t}$ is
the spectral resolution for \emph{A}.

The problem at hand is to express solutions of

\begin{equation*}
\tilde{A}f=Ef,\text{ \ \ \ \ }for\text{ }f\in D(\tilde{A}),\text{ }E\in \rho
(A)\cap \mathbb{R}
\end{equation*}
by eigenvectors of $U_{E0}\tilde{B}^{-1}.$ This is given by the following
theorem.

\begin{theorem}
A pair $f\in \mathcal{H},$ $E\in \rho (A)\cap \mathbb{R}$ solves (\ref{wy0})
if and only if there exists $g\in \mathcal{K}_{E}$ so that 
\begin{equation}
f=\tilde{A}^{-1}g  \label{pom2}
\end{equation}
and condition 
\begin{equation}
a_{g}(E)=E^{-1}.  \label{pom3}
\end{equation}
holds.
\end{theorem}

\begin{proof}
Observe that $g\in \ker (U_{0E}-E\tilde{B}^{-1})$ iff $g\in \mathcal{K}_{E}$
and $a_{g}(E)=E^{-1}$. Then, by theorem 2 we get that $f\in \mathcal{H},$ $%
E\in \rho (A)\cap \mathbb{R}$ solve (\ref{wy0}$)$ iff $f$ is given by (\ref
{pom2}) where $g\in \mathcal{K}_{E}$ and condition (\ref{pom3}) is satisfied.
\end{proof}

Now, we shall be interested in eigenvalues problem for $E\in \sigma (A).$
Recall that we assume $\sigma (A)=\sigma _{ac}(A)$.

For further discussion we shall need the spectral representation of $A$
supplied by the following theorem.

Put $\mathbb{I}$ for a countable set.

\begin{theorem}[Spectral Theorem]
Let A be self-adjoint in separable Hilbert space $\mathcal{H}$. Then there
exist measures $\{\mu _{n}\}_{n\in \mathbb{I}}$ on $\sigma (A)$ and unitary
operator U:$\mathcal{H\rightarrow }\underset{n\in \mathbb{I}}{\oplus }L^{2}(%
\mathbb{R},d\mu _{n})$ so that 
\begin{equation*}
(UAU^{-1}\psi )_{n}(t)=t\psi _{n}(t)
\end{equation*}
where $\psi $ is a vector from $\underset{n\in \mathbb{I}}{\oplus }L^{2}(%
\mathbb{R},d\mu _{n})$ with coordinates $\psi _{n}\in L^{2}(\mathbb{R},d\mu
_{n})$ for each $n\in \mathbb{I}.$
\end{theorem}

By the unitarity of $U$ the direct sum decomposition $\underset{n\in \mathbb{%
I}}{\oplus }L^{2}(\mathbb{R},d\mu _{n})$ can be ''transmitted'' to $\mathcal{%
H}.$\ Precisely, we have 
\begin{equation}
\mathcal{H}=\underset{n\in \mathbb{I}}{\oplus }\mathcal{H}^{n}\text{\ \ \ \ 
\emph{where }}\mathcal{H}^{n}=U^{-1}L^{2}(\mathbb{R},d\mu _{n}).
\label{decoHn}
\end{equation}
For all $n\in \mathbb{I}$ introduce 
\begin{equation}
A^{n}=A\lfloor \mathcal{H}^{n}\text{ }\emph{and}\text{ }\mathbb{A}^{n}=%
\mathbb{A}\lfloor \mathcal{H}^{n}.  \label{decoAn}
\end{equation}
Clearly $\sigma (A^{n})=supp\mu _{n}.$

For further discussion we need the following definition.

\begin{description}
\item[\emph{Definition}]  \emph{Let }$\widetilde{A}\in \mathcal{A}_{s}(A)$%
\emph{\ and }$\tilde{B}^{-1}=\widetilde{A}^{-1}-A^{-1}.$\emph{\ We will say
that operator }$\widetilde{A}$\emph{\ satisfies condition }$\mathbf{\hat{%
\sigma}}$ \emph{if for all n}$\in \mathbb{I}$\emph{\ we have}
\end{description}

\begin{equation}
Ran\tilde{B}^{-1}\cap Ran(A^{n}-E)=\{0\}\ \ \ \ \emph{where\ E\ runs\ over}%
\text{ }\sigma (A^{n}).  \label{ass}
\end{equation}

In the next section we will discuss condition (\ref{ass}) in more details.

Let $\tau $\emph{\ }be a subset of $\mathbb{I}$. Define 
\begin{equation}
\mathcal{H}^{\tau }:=\underset{j=1}{\oplus }\mathcal{H}^{j}  \label{decomH}
\end{equation}
and 
\begin{equation}
A^{\tau }=A\lfloor \mathcal{H}^{\tau }.  \label{decomA}
\end{equation}
For $E\in \rho (A^{\tau })\cap \mathbb{R}$ put $U_{E0}^{\tau }\equiv A^{\tau
}(A^{\tau }-E)^{-1},$\ $U_{0E}^{\tau }\equiv (U_{E0}^{\tau })^{-1}$\ and $%
\mathcal{K}_{E}^{\tau }=\{g\in \mathcal{H}^{\tau }$\emph{\ and g is an
eigenvector of }$U_{E0}^{\tau }\tilde{B}^{-1}\}.$

Now we are in position to formulate the following theorem.

\begin{theorem}
Assume that $\widetilde{A}\in \mathcal{A}_{s}(A)$\ satisfies $\mathbf{\hat{%
\sigma}}$ and g$\in \mathcal{H}^{\tau }.$\ A pair $f=\tilde{A}^{-1}g,$ $E\in
\sigma (A)$ solves 
\begin{equation}
\tilde{A}f=Ef  \label{wys}
\end{equation}
iff $E\in \varrho (A^{\tau })\cap \sigma (A),$ g$\in \mathcal{K}_{E}^{\tau }$%
\ and condition\ 
\begin{equation}
a_{g}^{\tau }(E)\equiv \left\| g\right\| ^{-2}(U_{E0}^{\tau }\tilde{B}%
^{-1}g,g)=E^{-1}  \label{fua}
\end{equation}
holds.
\end{theorem}

\begin{proof}
Let $\widetilde{A}\in \mathcal{A}_{s}(A)$ satisfies $\mathbf{\hat{\sigma}}$
and $g\in \mathcal{H}^{\tau }.$ Assume that a pair $f=\tilde{A}^{-1}g,$\ $%
E\in \sigma (A)$\ solve (\ref{wys}). Using $\widetilde{A}^{-1}=A^{-1}+\tilde{%
B}^{-1}$\ we can rewrite (\ref{wys}) in the form 
\begin{equation*}
E^{-1}g=(A^{-1}+\tilde{B}^{-1})g.
\end{equation*}
A direct calculation yields 
\begin{equation}
(A-E)A^{-1}g=E\tilde{B}^{-1}g.  \label{po3}
\end{equation}
Then $(A-E)A^{-1}g\in Ran(A^{\tau }-E)\cap Ran\tilde{B}^{-1}.$\ Due to
condition (\ref{ass}) we come to the conclusion that $E\in \varrho (A^{\tau
}).$\ Moreover from (\ref{po3}) follows $g\in \mathcal{K}_{E}^{\tau }$\ and\ 
$a_{g}^{\tau }(E)=E^{-1}.$

Conversely, assume $E\in \varrho (A^{\tau })\cap \sigma (A)$\ and $g\in 
\mathcal{H}^{\tau }.$\ Besides, let\ $g\in \mathcal{K}_{E}^{\tau }$\ and
condition (\ref{fua}) holds. Then we have 
\begin{equation*}
(\mathbb{I}-EU_{E0}^{\tau }\tilde{B}^{-1})g=0
\end{equation*}
or equivalently 
\begin{equation}
(U_{0E}^{\tau }-E\tilde{B}^{-1})g=0.  \label{po4}
\end{equation}
In turn (\ref{po4}) can be written as 
\begin{equation*}
A^{-1}g+\tilde{B}^{-1}g=E^{-1}g.
\end{equation*}
Putting $f=A^{-1}g+\tilde{B}^{-1}g=\tilde{A}^{-1}g$\ we come to (\ref{wys}).
\end{proof}

Theorems 3 and 5 solve the eigenvalues problem for operator $\tilde{A}\in 
\mathcal{A}_{s}(A)$ satisfying condition $\mathbf{\hat{\sigma}.}$ Moreover,
from above mentioned theorems we immediately get.

\begin{corollary}
Let $\tilde{A}\in \mathcal{A}_{s}(A)$ and satisfy $\mathbf{\hat{\sigma}.}$
If E$\in \sigma _{p}(\tilde{A})$ then E$\in \rho (A)\cap \mathbb{R}.$
\end{corollary}

Particularly, if $\mathbb{I}=\{1\}$ and $E\in \sigma _{p}(\tilde{A})$ then $%
E\in \rho (A)\cap \mathbb{R}.$

We will formulate our next results for the following case. Let again $\emph{A%
}$ be a self-adjoint, strictly positive operator in $\mathcal{H}$ with lower
bound $\lambda .$ Additionally we assume that \emph{A} has the following
orthogonal sum decomposition 
\begin{equation*}
A=\oplus _{n\in \mathbb{I}}A^{n}
\end{equation*}
where $A^{n}$ are operators defined by (\ref{decoAn}) with spectrum sets 
\begin{equation}
\sigma (A^{n})=[m_{n},\infty ),\text{ \ \ }m_{n}\geq \lambda .
\label{specAn}
\end{equation}
In fact, it is not difficult to generalize further results for the case when 
$A^{n}$ have absolutely continuous spectrums with gaps. However, for
simplicity we make assumption (\ref{specAn}).

Define operator $\tilde{B}^{-1}\in \mathcal{B}(A)$\ by 
\begin{equation}
\tilde{B}^{-1}=\sum_{k\in \mathbb{S}}b_{k}^{-1}(\cdot ,e_{k})e_{k}
\label{invari}
\end{equation}
where $b_{k}$\ are real constants, $b_{k}\neq 0$, $\{e_{k}\}_{k\in \mathbb{S}%
}$\ is an orthogonal basis in $N_{0}$\ and for each $k\in \mathbb{S}$ there
exists $\tau _{k}\subset \mathbb{I},$ so that \emph{e}$_{k}\in \mathcal{H}%
^{\tau _{k}}$ (see \ref{decomH}) and $\tbigcap\limits_{k\in \mathbb{S}}\tau
_{k}=\{0\}.$\ 

Now, we will study the eigenvalues problem for $\tilde{A}\in \mathcal{A}%
_{s}(A)$ where $\tilde{B}^{-1}=\tilde{A}^{-1}-A^{-1}$ has the form (\ref
{invari}).\ In particular, our aim will be to characterize the following
functions 
\begin{equation*}
N_{-}(\tilde{A})=\#\{E<0:E\text{\emph{solves} }(\ref{wys})\}
\end{equation*}
and 
\begin{equation*}
N_{+}(\tilde{A})=\#\{E>0:E\text{\emph{solves} }(\ref{wys})\}
\end{equation*}
where \# denotes the number of elements of the set $\{\cdot \}$ including
its multiplicity.

One can easily note that 
\begin{equation*}
\sigma (A^{\tau _{k}})=[M_{k},\infty )\text{ \ \ \ \ \emph{where }}%
M_{k}\equiv \inf_{n\in \tau _{k}}m_{n}.
\end{equation*}
Let $E_{k}\in \mathbb{R}\backslash \lbrack M_{k},\infty ).$ Henceforth, we
will omit the subscript \emph{k} putting $E_{k}\equiv E$. We also abbreviate 
$\mathcal{K}_{E}^{k}\equiv \mathcal{K}_{E}^{\tau _{k}},$\ $U_{E0}^{k}\equiv
U_{E0}^{\tau _{k}}$\ and $U_{0E}^{k}\equiv U_{0E}^{\tau _{k}}.$

Assume $g\in \mathcal{K}_{E}^{k}.$ Then we have 
\begin{equation*}
U_{E0}^{k}\tilde{B}^{-1}g=b_{k}^{-1}(g,e_{k})U_{E0}^{k}e_{k}.
\end{equation*}
Since \emph{g} is an eigenvalue of $U_{E0}^{k}\tilde{B}^{-1}$ we conclude
that \emph{g} belongs to the subspace spanned by $U_{E0}^{k}e_{k}.$\ For
simplicity we put $g=U_{E0}^{k}e_{k}.$\ With this observation theorems 3 and
5 read.

\begin{theorem}
Let \~{A} satisfy $\mathbf{\hat{\sigma}}$ and \~{B}$^{-1}$\ be defined by (%
\ref{invari}). Assume that given k$\in \mathbb{S}$ there exists number E$\in 
\mathbb{R}$\TEXTsymbol{\backslash}\{0\} so that condition 
\begin{equation}
a_{k}(E)\equiv b_{k}^{-1}(U_{E0}^{k}e_{k},e_{k})=E^{-1}  \label{agk}
\end{equation}
fulfills. Then E$\in \mathbb{R}\backslash \lbrack M_{k},\infty ),$ and the
pair E, \ f=$\tilde{A}^{-1}g_{k}$ where $g_{k}=U_{E0}^{k}e_{k}$ solves (\ref
{wys}). Moreover $\{f_{k}\}_{k\in \mathbb{S}^{\prime }}$ where $\mathbb{S}%
^{\prime }=\{k\in \mathbb{S}$:(\ref{agk}) holds$\}$ is the complete system
of eigenvectors for \~{A}.
\end{theorem}

From above theorem follows that $E\in \sigma _{p}(\tilde{A})$ iff $E$ solves
(\ref{agk}).

For each $\emph{k}\in \mathbb{S}$ introduce functions on $\mathbb{R}%
\backslash \lbrack M_{k},\infty )$\ defined by 
\begin{equation}
s_{k}(E)\equiv E(U_{E0}^{k}e_{k},e_{k})=E\int_{[M_{k},\infty )}\frac{t}{t-E}%
d\nu _{k}(t)  \label{reprsk}
\end{equation}
where $\nu _{k}(t)$ stands for the spectral measure associated to $\emph{e}%
_{k}$\ \emph{i.e.} $\nu _{k}(t)=(\mathbb{E}_{t}e_{k},e_{k}).$\ With this
notation (\ref{agk}) can be rewritten as 
\begin{equation}
s_{k}(E)=b_{k}.  \label{sk}
\end{equation}
Obviously, by the construction $s_{k}(E)$\ are continuous. Moreover the fact
that $\frac{Et}{t-E}$\ grows monotonically as the function of \emph{E}
implies that $s_{k}(E)$ grow monotonically too. Besides, $E<0$\ iff $%
s_{k}(E)<0.$\ Then as follows from (\ref{sk}) \emph{E}$\in \sigma _{p}(%
\tilde{A})$\ iff 
\begin{equation*}
\emph{b}_{k}\in (s_{k}(-\infty ),s_{k}(M_{k}))\ 
\end{equation*}
where $s_{k}(-\infty )\equiv \lim_{E\rightarrow \infty }s_{k}(E)$\ and $%
s_{k}(M_{k})\equiv \lim_{E\rightarrow M_{k}}s_{k}(E).$\ So, we have to find $%
s_{k}(-\infty )$ and $s_{k}(M_{k}).$\ The first result is given in the
following statement.

\begin{proposition}
For e$_{k}\in \mathcal{H}_{1}$ the expression $s_{k}(-\infty )$ is finite
and given by 
\begin{equation*}
s_{k}(-\infty )=-\left\| e_{k}\right\| _{1}^{2}.
\end{equation*}
Otherwise, i.e. for e$_{k}\in \mathcal{H}_{0}\backslash \mathcal{H}_{1}$ we
have $s_{k}(-\infty )=-\infty .$
\end{proposition}

\begin{proof}
From (\ref{reprsk}) we immediately get that $s_{k}(-\infty )=-\left\langle 
\mathbb{A}e_{k},e_{k}\right\rangle =-\left\| e_{k}\right\| _{1}^{2}.$
Clearly, this expression is finite iff $e_{k}\in \mathcal{H}_{1}.$
\end{proof}

To present the next result we need some preparation. For $i=1,$ $2$ define
space $\mathcal{H}_{q}^{\prime }$\ as the completion of \emph{D(A)}\ in the
norm 
\begin{equation}
\left\| u\right\| _{i}^{\prime }=\left\| \sum_{k\in \mathbb{S}%
}(A^{k}-M_{k})^{i/2}u\right\| _{0}.  \label{homosp}
\end{equation}
Let $\mathcal{H}_{-i}^{\prime }$ stand for the dual spaces to $\mathcal{H}%
_{i}^{\prime }$ with the respect to $\mathcal{H}$. We keep \emph{q} for the
integer with $\left| q\right| \leq 2.$ Note, that $\mathcal{H}_{q}^{\prime }$%
\ in some sense generalize the notation of a homogenous Sobolev space \cite
{Ma}. In further discussion we shall use only $\mathcal{H}_{1}^{\prime }$\
and $\mathcal{H}_{-1}^{\prime }.$\ Obviously, we have $\mathcal{H}%
_{1}\subset \mathcal{H}_{1}^{\prime }$\ and $\mathcal{H}_{-1}^{\prime
}\subset \mathcal{H}_{-1}.$

\begin{proposition}
For $\mathbb{A}^{1/2}e_{k}\in \mathcal{H}_{-1}^{\prime }$ the expression\ $%
s_{k}(M_{k})$ is finite and given by 
\begin{equation}
s_{k}(M_{k})=M_{k}\left\| \mathbb{A}^{1/2}e_{k}\right\| _{-1}^{\prime 2}.
\label{auxske}
\end{equation}
Otherwise, i.e. for $\mathbb{A}^{1/2}e_{k}\in \mathcal{H}_{-1}\backslash 
\mathcal{H}_{-1}^{\prime }$\ we have $s_{k}(M_{k})=\infty .$\ \ \ 
\end{proposition}

\begin{proof}
Note that $s_{k}(E)=E(U_{E0}^{k}e_{k},e_{k})$ can be equivalently written as 
$s_{k}(E)=E\left\| A^{1/2}(A-E)^{-1/2}e_{k}\right\| ^{2}.$\ Then we have $%
s_{k}(M_{k})=M_{k}\left\| \mathbb{A}^{1/2}e_{k}\right\| _{-1}^{\prime 2}$\
which is finite iff $\mathbb{A}^{1/2}e_{k}\in \mathcal{H}_{-1}^{\prime }.$
\end{proof}

To summarize above discussion let us select four different cases.

\QTP{Body Math}
$Case$\textbf{\ }$1.$ Let $e_{k}\in \mathcal{H}_{0}\backslash \mathcal{H}%
_{1} $ and $\mathbb{A}^{1/2}e_{k}\in \mathcal{H}_{-1}\backslash \mathcal{H}%
_{-1}^{\prime }$ for all $k\in \mathbb{S}$. Then 
\begin{equation*}
N_{-}=\#\{b_{k}:b_{k}<0\},\text{ \ \ }N_{+}=\#\{b_{k}:b_{k}>0\}.
\end{equation*}

\QTP{Body Math}
$Case$\textbf{\ }$2.$ Let $e_{k}\in \mathcal{H}_{0}\backslash \mathcal{H}%
_{1} $ and $\mathbb{A}^{1/2}e_{k}\in \mathcal{H}_{-1}^{\prime }$ for all $%
k\in \mathbb{S}$. Then 
\begin{equation*}
N_{-}=\#\{b_{k}:b_{k}<0\},\text{ \ \ }N_{+}=\#\{b_{k}:0<b_{k}\leq
M_{k}\left\| \mathbb{A}^{1/2}e_{k}\right\| _{-1}^{\prime 2}\}.
\end{equation*}

\QTP{Body Math}
$Case$\textbf{\ }$3.$ Let $e_{k}\in \mathcal{H}_{1}$ and\ $\mathbb{A}%
^{1/2}e_{k}\in \mathcal{H}_{-1}\backslash \mathcal{H}_{-1}^{\prime }$\ for
all $k\in \mathbb{S}$. Then 
\begin{equation*}
N_{-}=\#\{b_{k}:-\left\| e_{k}\right\| _{1}^{2}\leq b_{k}<0\},\text{ \ }%
N_{+}=\#\{b_{k}:b_{k}>0\}.
\end{equation*}

\QTP{Body Math}
$Case$\textbf{\ }$4.$\textbf{\ }Let $e_{k}\in \mathcal{H}_{1}$ and\ $\mathbb{%
A}^{1/2}e_{k}\in \mathcal{H}_{-1}^{\prime }$\ for all $k\in \mathbb{S}$.
Then 
\begin{equation*}
N_{-}=\#\{b_{k}:-\left\| e_{k}\right\| _{1}^{2}\leq b_{k}<0\},\text{ \ }%
N_{+}=\#\{b_{k}:0<b_{k}\leq M_{k}\left\| \mathbb{A}^{1/2}e_{k}\right\|
_{-1}^{\prime 2}\}.
\end{equation*}

Observe that for cases 1, 2 operator $\tilde{A}$ belongs to strongly
singular class $\mathcal{A}_{ss}(A)$. In turn, for cases 3, 4 $\tilde{A}$
belongs to weakly singular class $\mathcal{A}_{ws}(A)$.

We also remark that for $e_{k}\in \mathcal{H}_{1}$ condition $\mathbb{A}%
^{1/2}e_{k}\in \mathcal{H}_{-1}^{\prime }$\ is equivalent to $\mathbb{A}%
e_{k}\in \mathcal{H}_{-1}^{\prime }.$\ Indeed, one can show 
\begin{equation*}
M_{k}\left\| \mathbb{A}^{1/2}e_{k}\right\| _{-1}^{\prime 2}=\left\| \mathbb{A%
}e_{k}\right\| _{-1}^{\prime 2}-\left\| e_{k}\right\| _{1}^{\prime 2}.
\end{equation*}
So, the pure point spectrum $\sigma _{p}(\tilde{A})$\ can be characterized
in the terms of $\mathbb{A}e_{k}$ as was done in \cite{AKK2} for the case $%
\tilde{A}\in \mathcal{A}_{ws}(A).$

Let us make short digression about the absolutely continuous spectrum of $%
\tilde{A}$ with $\tilde{B}^{-1}=\tilde{A}^{-1}-A^{-1}$ of type (\ref{invari}%
). Note that we have 
\begin{equation*}
\sigma _{ac}(\tilde{A})=\tbigcup_{k\in \mathbb{S}}\sigma (\tilde{A}^{\tau
_{k}})
\end{equation*}
where the inverse of $\tilde{A}^{\tau _{k}}$ has the form 
\begin{equation*}
(\tilde{A}^{\tau _{k}})^{-1}=(A^{\tau _{k}})^{-1}+b_{k}^{-1}(\cdot
,e_{k})e_{k}.
\end{equation*}
So, we see that $\sigma _{ac}(\tilde{A}^{\tau _{k}})=\sigma _{ac}(A^{\tau
_{k}})=[M_{k},\infty ).$ Moreover, observing that $\tbigcup\limits_{k\in 
\mathbb{S}}[M_{k},\infty )=[\lambda ,\infty )$ we get

\begin{equation}
\sigma _{ac}(\tilde{A})=\sigma (A)=[\lambda ,\infty ).  \label{abcosp}
\end{equation}

\section{EIGENVALUES PROBLEM FOR ADDITIVE STRONGLY SINGULAR PERTURBATION OF
LAPLACE OPERATOR.}

In this section we put $\mathcal{H}=L^{2}(\mathbb{R}^{3},dx)\equiv L^{2}$
and $A=-\Delta +\lambda :D(A)\rightarrow L^{2}$\ where $\Delta $ stands for
the self-adjoint Laplace operator in $L^{2}$\ and $\lambda >0.$ Clearly, we
have $\sigma (A)=[\lambda ,\infty ).$ For our convenience we put $G:=A^{-1}.$
Then $G$ is an integral operator with kernel $G(x-y)$ given by 
\begin{equation}
G(x)=\frac{1}{4\pi }\frac{\exp (-\sqrt{\lambda }\left| x\right| )}{\left|
x\right| }.  \label{G(x)}
\end{equation}

As in general discussion we construct the chain (\ref{cha}) of Hilbert
spaces. Now, $\mathcal{H}_{q}$\ defined as the completions of $D(A)$ in
norms 
\begin{equation}
\left\| f\right\| _{q}^{2}=\int_{\mathbb{R}^{3}}\left| (-\Delta +\lambda
)^{q/2}f(x)\right| ^{2}dx  \label{norlap}
\end{equation}
\ coincides with the Sobolev spaces $W^{2,q}(\mathbb{R}^{3})\equiv W^{2,q}.$
As before, we put $\mathbb{A}$\ for the extension of $A:D(A)\subset
L^{2}\rightarrow W^{2,-2}$\ and $\mathbb{G}$ for its inverse.

Consider operator $\mathbb{V}:D(\mathbb{V)\subset }C(\mathbb{R}^{3})\equiv
C^{0}\rightarrow W^{2,-2}$\ satisfying two conditions

\begin{itemize}
\item  $K)$ $ker\mathbb{V\cap }W^{2,2}$ is dense in $L^{2}.$

\item  $R)$ $Ran\mathbb{V\mathbb{\cap }}W^{2,-1}=\{0\}.$
\end{itemize}

The technics proposed in \cite{SK1} allows to construct the operator
belonging to $\mathcal{A}_{s}(A)$ which\ is, in some sense, a sum of $%
\mathbb{A}$\ and $\mathbb{V}$.\ The concept is based on the analogy with the
generalized sum (see for example \cite{AlKo,Ber,TKVK,VK3,KrYa}). Now, we
present the main results of \cite{SK1}.

Let us introduce $G_{r},$\ $G_{s}$\ for the integral operator with kernels
given by 
\begin{equation*}
G_{s}(x):=\frac{1}{4\pi }\frac{1}{\left| x\right| },\text{\ \ }%
G_{r}(x):=G(x)-G_{s}(x)\text{.}
\end{equation*}
Define 
\begin{equation*}
C_{r}:=\mathbb{I}+G_{r}\mathbb{V}:D(C_{r})=\{g\in D(\mathbb{V}):C_{r}g\in
W^{2,2}\}\rightarrow W^{2,2}
\end{equation*}
and 
\begin{equation*}
C_{s}:=\mathbb{I}-G_{s}\mathbb{V}:D(C_{s})=D(C_{r})\rightarrow L^{2}.
\end{equation*}
Assume that $C_{r}$ is invertible.\ Then one can show the invertibility of $%
C_{s}.$ Indeed, let $C_{s}h=0.$ Then $C_{r}-G\mathbb{V}h=0.$ Due to \emph{R)}
and the fact that $\ker C_{s}=\{0\}$ we get \emph{h=0}.

Let $f\in D(C_{s}^{-1})$ and $f_{r}=C_{s}^{-1}f.$\ Define the set 
\begin{equation}
D(\mathbb{A\hat{+}V)=\{}f\in D(C_{s}^{-1}):\mathbb{A}f\mathbb{+V}f_{r}\in
L^{2}\}  \label{dsu}
\end{equation}
and operator $\mathbb{A\hat{+}V}$ which acts as 
\begin{equation}
(\mathbb{A\hat{+}V)}f=\mathbb{A}f\mathbb{+V}f_{r},\text{ \ \ \ }f\in D(%
\mathbb{A\hat{+}V)}.  \label{sum}
\end{equation}
To\ explain (\ref{dsu}) and (\ref{sum}) we assume $f\in D(\mathbb{A\hat{+}V)}
$. Then $\mathbb{A}f\mathbb{+V}f_{r}=g\in L^{2}$\ implies $f=Gg-\mathbb{GV}%
f_{r}.$\ Since $Gg\in W^{2,2}$\ by the Sobolev theorem we have $Gg\in C^{0}.$%
\ However $f$ possess singularity induced by $\mathbb{GV}f_{r}.$\ Thus, to
regularize $f$ consider $f+G_{s}\mathbb{V}f_{r}$. On the other hand $%
f_{r}=C_{s}^{-1}f$\ yields $f_{r}=f+G_{s}\mathbb{V}f_{r}.$\ So we see that $%
f_{r}$\ is just a regularization of $f$.

With this notation we have the following theorem.

\begin{theorem}
Let us assume that $C_{r}$ is invertible and operator 
\begin{equation}
\tilde{B}^{-1}=-\mathbb{GV}C_{r}^{-1}G:D(\tilde{B}^{-1})=AD(C_{r}^{-1})%
\rightarrow L^{2}  \label{bsum}
\end{equation}
is self-adjoint. Then $\mathbb{A\hat{+}V\in }\mathcal{A}_{s}(A)$ and its
inverse is given by 
\begin{equation}
(\mathbb{A\hat{+}V)}^{-1}=A^{-1}+\tilde{B}^{-1}:D(\tilde{B}^{-1})\rightarrow
L^{2}.  \label{kkosum}
\end{equation}
\ \ \ \ 
\end{theorem}

\begin{proof}
Let$\ C_{r}$ be invertible. First we shall show that $A\widehat{+}\mathbb{V}$
is invertible also and its inverse has the form (\ref{kkosum}). Let $f\in
\ker (A\widehat{+}\mathbb{V)}$ and $f_{r}=C_{s}^{-1}f.$ Then 
\begin{equation*}
\mathbb{A}f+\mathbb{V}f_{r}=0
\end{equation*}
i.e. 
\begin{equation*}
f=-\mathbb{GV}f_{r}=-\mathbb{GV}C_{s}^{-1}f.
\end{equation*}
Then we get 
\begin{equation*}
C_{r}C_{s}^{-1}f=0.
\end{equation*}
Since $\ker C_{r}=\ker C_{s}^{-1}=\{0\}$ we obtain $f=0.$ Now let $f\in D(A%
\widehat{+}\mathbb{V}),$ $f_{r}=C_{s}^{-1}f$. Then 
\begin{equation*}
g=(A\widehat{+}\mathbb{V)}f=\mathbb{A}f+\mathbb{V}f_{r}\in L^{2}.
\end{equation*}
So, we have 
\begin{equation}
f=A^{-1}g-\mathbb{GV}C_{s}^{-1}f.  \label{tec251}
\end{equation}
From \ref{tec251} follows 
\begin{equation}
C_{r}C_{s}^{-1}f=Gg,  \label{tec252}
\end{equation}
i.e. $Gg\in D(C_{r}^{-1}).$ Finally after inserting (\ref{tec252}) to (\ref
{tec251}) we get 
\begin{equation*}
f=A^{-1}g-\mathbb{GV}C_{r}^{-1}Gg=A^{-1}g+\tilde{B}^{-1}g.
\end{equation*}
This means that operator $(A\widehat{+}\mathbb{V})^{-1}$ is given by (\ref
{kkosum}). Now we shall show that $\tilde{B}^{-1}\in \mathcal{B}(A).$ For
this aim let us note that by a self-adjointness of $\tilde{B}^{-1}$ we have $%
G\ker \tilde{B}^{-1}=[GRan\tilde{B}^{-1}]^{\perp }$, where $\perp $ denotes
orthogonal completion in $W^{2,2}$ topology. Further let us note that $[%
\mathbb{G}^{2}RanV]^{\perp }\subseteq \lbrack GRan\tilde{B}^{-1}]^{\perp }.$
In turn, since operator $\mathbb{V}$ has property $R)$ we get 
\begin{equation*}
\mathbb{G}Ran\mathbb{V}\cap W^{2,2}=\{0\}.
\end{equation*}
Using theorem A.1 \cite{AKK} we can conclude that $[\mathbb{G}%
^{2}RanV]^{\perp }$ is dense in $L^{2}.$ Hence we get the density of $[%
\mathbb{G}^{2}RanV]^{\perp }$ and $[GRan\tilde{B}^{-1}]^{\perp }$ in $L^{2}.$
Operator $\tilde{B}^{-1}$ belongs to $\mathcal{B}(A).$ Then by theorem 1 we
get $A\widehat{+}\mathbb{V}\in \mathcal{A}_{ss}(A).$
\end{proof}

In fact the above construction of $A\widehat{+}\mathbb{V}$ can be repeated
for an arbitrary $d\geq 3.$ However in general case we define $G_{s}$ as the
integral operator with the kernel $G_{s}(x)=c\frac{1}{\left| x\right| ^{d-2}}
$ where $c$ is an apropriated constance.

To investigate eigenvalues problem for $\mathbb{A\hat{+}V}$ let us recall
spectral representation of $A:=-\Delta +\lambda :D(A)\rightarrow L^{2}.$ Put 
$r,$ $\theta ,\phi $ for the spherical coordinates in $\mathbb{R}^{3}.$ It
is known that $L^{2}(\mathbb{R}^{3})$ can be written as 
\begin{equation*}
L^{2}(\mathbb{R}^{3})=L^{2}([0,\infty ),r^{2}dr)\otimes L^{2}(S^{1})
\end{equation*}
where $S^{1}$ is unit sphere. This leads to the following direct sum
decomposition

\begin{equation}
L^{2}(\mathbb{R}^{3})=\overset{\infty }{\underset{l=0}{\oplus }}\overset{l}{%
\underset{m=-l}{\oplus }}\mathcal{H}^{lk},\text{ \ \ \ \ \ \ \ }\mathcal{H}%
^{lk}=L^{2}([0,\infty ),r^{2}dr)\otimes Y_{lk}(\theta ,\phi )  \label{deL2R3}
\end{equation}
where $Y_{lk}(\theta ,\phi )$ are spherical harmonics. Then $f\in L^{2}$ can
be represented in the following form

\begin{equation*}
f(x)=\sum\limits_{l=o}^{\infty
}\sum\limits_{k=-l}^{l}\int\limits_{0}^{\infty }dpp^{2}\tilde{f}%
_{lk}(p)j_{l}(pr)Y_{lk}(\theta ,\phi )
\end{equation*}
where $j_{l}(pr)$ are spherical Bessel's functions. Keeping consistency with
(\ref{deL2R3}) define unitary operator

\begin{equation}
U:L^{2}(\mathbb{R}^{3})\rightarrow \overset{\infty }{\underset{l=0}{\oplus }}%
\overset{l}{\underset{k=-l}{\oplus }}L^{2}([\lambda ,\infty ),d\mu _{lk}),%
\text{ \ \ \ \ }d\mu _{lk}(t)=\frac{1}{2}\sqrt{t-\lambda }dt  \label{operaU}
\end{equation}
by 
\begin{equation*}
(Uf)_{lk}=\tilde{f}_{lk}(\sqrt{p-\lambda }).
\end{equation*}
Then we have 
\begin{equation*}
U(-\Delta +\lambda )U^{-1}\psi )_{lk}=t\psi _{lk}(t),\text{ \ }\psi \in 
\overset{\infty }{\underset{l=0}{\oplus }}\overset{l}{\underset{k=-l}{\oplus 
}}L^{2}([\lambda ,\infty ),d\mu _{lk}).
\end{equation*}
In accordance with (\ref{deL2R3}) operator \emph{A} can be decomposed as 
\begin{equation*}
A=\overset{\infty }{\underset{l=0}{\oplus }}\overset{l}{\underset{k=-l}{%
\oplus }}A^{lk};\ \ A^{lk}=A\lfloor \mathcal{H}^{lk}
\end{equation*}
and 
\begin{equation}
\sigma (A^{lk})=[\lambda ,\infty )\text{ \ \ \emph{for all }}\emph{l,\ k}.
\label{spectA}
\end{equation}

Assume that $\tilde{A}\in \mathcal{A}_{s}(A)$ and its inverse is given by $%
\tilde{A}^{-1}=A^{-1}+\tilde{B}^{-1}$ where $A=-\Delta +\lambda .$ Observe
that by (\ref{spectA}) $\tilde{A}$ satisfies $\mathbf{\hat{\sigma}}$ (\ref
{ass}) iff 
\begin{equation}
Ran\tilde{B}^{-1}\cap Ran(A-E)=\{0\}\text{ \ \ \ \emph{for all }}\emph{E}\in
\lbrack \lambda ,\infty ).  \label{assA}
\end{equation}
The problem at hand is to select class of operators $\mathbb{V}$ so that $A%
\hat{+}\mathbb{V}$ satisfy $\mathbf{\hat{\sigma}}.$

Let $\emph{N}$ be a compact with boundary of $C^{1}$ class. We shall say
that operator $\mathbb{V}$ satisfying K), R) has the property $\mathbf{\hat{N%
}}$\ if any $\mu \in Ran\mathbb{V}$ is a distribution from $W^{2,2}$
supported by the set $N_{\mu }\subset N$.

\begin{theorem}
Let $\mathbb{V}$ have property \textbf{\^{N}} and $\mathbb{A\hat{+}V\in }%
\mathcal{A}_{s}(A).$ Then $\mathbb{A\hat{+}V}$\ satisfies $\mathbf{\hat{%
\sigma}}$.
\end{theorem}

\begin{proof}
Let $\mathbb{V}$ have property \textbf{\^{N}}. Assume that $g\in Ran\tilde{B}%
^{-1}\cap Ran(A-E)$ where $E\in \lbrack \lambda ,\infty ).$ Since $Ran\tilde{%
B}^{-1}\subseteq Ran\mathbb{G}\mathbb{V}$ (see (\ref{bsum}))\ there exists $%
\mu \in W^{2,-2}$ supported by the compact set $N_{\mu }$ so that $g=\mathbb{%
G}\mu .$\ Clearly, $\mu \in Ran(\mathbb{A}-E).$\ Let $f\in \mathcal{H}$
satisfies 
\begin{equation}
\mu (x)=(\mathbb{A}-E)f(x).  \label{mif}
\end{equation}
Expressing $G(x-y)$ (\ref{G(x)}) in the spherical system of coordinates one
can show that $\left| \mathbb{G}\mu \right| $ behaves like $\frac{e^{-\left|
x\right| }}{\left| x\right| }$ as $\left| x\right| \rightarrow \infty .$
Thus $g\in L^{2,s}\equiv L^{2}(\mathbb{R}^{3},(\left| x\right| ^{2}+\lambda
)^{s}dx),$ $s>1/2$ and $\mu \in W^{2,-2,s}\equiv W^{2,-2}((\mathbb{R}%
^{3},(\left| x\right| ^{2}+\lambda )^{s}dx).$ Relying on the results of \cite
{ABK} we get that for all $E\in \lbrack \lambda ,\infty ),$ $s>1/2$ the
limits 
\begin{equation*}
K^{\pm }\equiv \lim_{\varepsilon \rightarrow 0}(A-E\pm i\varepsilon
):W^{2,-2,s}\rightarrow L^{2,-s}
\end{equation*}
exist in the uniform operator norm. Besides 
\begin{equation*}
K^{\pm }\mu =f.
\end{equation*}
It known that $K^{\pm }$ can be represented as the integral operator with
the kernel given by 
\begin{equation*}
K^{\pm }(x-y)=(4\pi )^{-1}\exp (\pm i\sqrt{E-\lambda }\left| x-y\right|
)\left| x-y\right| ^{-1}.
\end{equation*}
Thefore $f$ can be written as 
\begin{equation*}
f(x)=\int_{\mathbb{R}^{3}}K^{\pm }(x-y)\mu (y)dy=\int_{N_{\mu }}K^{\pm
}(x-y)\mu (y)dy.
\end{equation*}
Similarly as before expressing $K^{\pm }(x-y)$ in the spherical system of
coordinates we get the following asymptotics $\left| f(x)\right| \sim \left|
x\right| ^{-1}$ for $\left| x\right| \rightarrow \infty .$ Then the fact
that $f\in L^{2}$ implies $f=0.$ Thus $\mu =0$ as well as $g=0$ and (\ref
{assA}) is satisfied.
\end{proof}

Above theorem shows that if $\mathbb{V}$ has property $\mathbf{\hat{N}}$
then operator $A\hat{+}\mathbb{V}$ satisfies $\mathbf{\hat{\sigma}}$ . This
result and the fact $\sigma (A^{lk})=\sigma (A)$ implies\ by corollary 6 the
following statement.\ 

\begin{proposition}
Let $\mathbb{V}$ have the property $\mathbf{\hat{N}}$ and $A\hat{+}\mathbb{%
V\in }\mathcal{A}_{s}(A).$ Then $\mathbb{\sigma }_{p}\mathbb{(}A\hat{+}%
\mathbb{V)}\subset (-\infty ,\lambda ).$
\end{proposition}

Now, we shall consider two particular cases of $\mathbb{A\hat{+}V}$ and
investigate eigenvalues problem relying on abstract results of section 2 and
3.

\section{EXAMPLES}

\subsection{STRONGLY SINGULAR PERTURBATION OF LAPLACE OPERATOR BY THE
DYNAMICS LIVING ON INTERVAL}

This example is not strictly in the scheme of our previous discussion
because we take $\mathcal{H}=L^{2}(\Omega ,dx)\equiv L^{2}(\Omega ),$ $%
\Omega =\left\langle 0,\pi \right\rangle \times \mathbb{R}^{2}$\ instead of $%
L^{2}(\mathbb{R}^{3},dx).$\ Let $A\equiv -\Delta $ stand for the Laplace
operator with the Dirichlet boundary condition\ in $L^{2}(\Omega )$%
\begin{equation*}
D(A)=\{\frac{1}{2}\pi ^{-3}\sum_{k=1}^{\infty }\int_{\mathbb{R}^{2}}d\text{%
\b{p}}\tilde{u}_{k}(\text{\b{p}})e^{i\text{\b{p}\b{x}}}\sin kx_{1};\text{ \ }%
\sum_{k=1}^{\infty }\int_{\mathbb{R}^{2}}d\text{\b{p}}\left| (p^{2}+k^{2})%
\tilde{u}_{k}(\text{\b{p}})\right| ^{2}<\infty \}\rightarrow L^{2}(\Omega ).
\end{equation*}
Then we have $\sigma _{ac}(A)=\sigma (A)=[1,\infty ).$ Setting $G$ for the
inverse of $A$ we obtain an integral operator with the kernel 
\begin{equation*}
G(x,y)=\frac{1}{2}\pi ^{-3}\sum_{k=1}^{\infty }\int_{\mathbb{R}^{2}}d\text{%
\b{p}}\frac{e^{i\text{\b{p}}(\text{\b{x}}-\text{\b{y}})}}{p^{2}+k^{2}}\sin
kx_{1}\sin ky_{1}
\end{equation*}
where \b{p}$=(p_{2},p_{3}),$ \b{x}$=(x_{2},x_{3})$\ and $%
p^{2}=p_{2}^{2}+p_{3}^{2}.$\ According to the general discussion (see (\ref
{cha})) we construct spaces $\mathcal{H}_{q}$ which are given by 
\begin{equation*}
\mathcal{H}_{q}=\{u(x)=\frac{1}{2}\pi ^{-3}\sum_{k=1}^{\infty }\int_{\mathbb{%
R}^{2}}d\text{\b{p}}\tilde{u}_{k}(\text{\b{p}})e^{i\text{\b{p}\b{x}}}\sin
kx_{1};\text{ \ }\sum_{k=1}^{\infty }\int_{\mathbb{R}^{2}}d\text{\b{p}}%
(p^{2}+k^{2})^{q}\left| \tilde{u}_{k}(\text{\b{p}})\right| ^{2}<\infty \}.
\end{equation*}
Now, our aim is to construct an operator from $\mathcal{A}_{s}(A)$
corresponding to the formal expression 
\begin{equation*}
-\Delta +\alpha (-\frac{\partial ^{2}}{\partial x_{1}^{2}})\delta (\text{\b{x%
}}).
\end{equation*}
where $\alpha \in \mathbb{R}$. Let$\ $%
\begin{equation*}
\mathbb{V}_{\alpha }\mathbb{\equiv -\alpha }\Delta _{1}\delta (\text{\b{x}}%
):D(\mathbb{V}_{\alpha }\mathbb{)}\rightarrow W^{2,-2}
\end{equation*}
\ be given by 
\begin{equation*}
\mathbb{V}_{\alpha }f=\sum_{k=1}^{\infty }\alpha k^{2}c_{k}(f)\sin
kx_{1}\delta (\text{\b{x}}),\text{ \ }D(\mathbb{V}_{\alpha }\mathbb{)=\{}f%
\mathbb{\in }C(\mathbb{\Omega }):\mathbb{V}_{\alpha }f\in W^{2,-2}\}
\end{equation*}
where $c_{k}(f)=\int_{0}^{\pi }dx_{1}f(x_{1},0,0)\sin kx_{1}$. It is not
hard to see that $Ran\mathbb{V}_{\alpha }\subset \mathcal{H}_{-2}\backslash 
\mathcal{H}_{-1}.$

Similarly as in the previous discussion we can construct operator $-\Delta 
\hat{+}\alpha (\mathbb{-}\Delta _{1})\delta ($\b{x}$)$ which acts $(-\Delta 
\hat{+}\alpha (\mathbb{-}\Delta _{1})\delta ($\b{x}$))f=-\Delta f+\alpha (%
\mathbb{-}\Delta _{1})\delta ($\b{x}$)f_{r}$\ where $f_{r}$ is some
regularization of\ $f.$\ This problem is discussed in detail in \cite{SK1}.
Now we give only the final result.

For each $k\in \mathbb{N}$ we $s_{k}=\pi ^{-2}(-\ln \frac{k}{2}+C)$ where $C$
is the Euler constant. We also define 
\begin{equation}
e_{k}(x):=\frac{1}{2}\pi ^{-3}\int_{\mathbb{\Omega }}dyG(x,y)\sin
ky_{1}\delta (\text{\b{y}})=\frac{1}{2}\pi ^{-3}\int_{\mathbb{R}^{2}}d\text{%
\b{p}}\frac{e^{i\text{\b{p}\b{x}}}}{(p^{2}+k^{2})}\sin kx_{1}.  \label{fuv}
\end{equation}
One can show by a direct calculation that $\{e_{k}\}_{k\in \mathbb{N}}$ is
the orthogonal system.

\begin{theorem}[\protect\cite{SK1}]
Operator $-\Delta \hat{+}(\mathbb{-}\Delta _{1})\delta ($\b{x}$)\in \mathcal{%
A}_{ss}(A)$ and its inverse is given by 
\begin{equation}
(-\Delta \hat{+}(\mathbb{-}\Delta _{1})\delta (\text{\b{x}}))^{-1}=G+\tilde{B%
}^{-1}:D(\tilde{B}^{-1})=\{f\in L^{2}(\Omega ):\tilde{B}^{-1}f\in
L^{2}(\Omega )\}  \label{spl1}
\end{equation}
where 
\begin{equation*}
\tilde{B}^{-1}f=\sum_{k=1}^{\infty }b_{k}^{-1}(f,e_{k})e_{k},\text{ \ }%
b_{k}^{-1}\equiv b_{\alpha ,k}^{-1}=-\alpha k^{2}(1+\alpha k^{2}s_{k})^{-1}.%
\text{\ }
\end{equation*}
\end{theorem}

To solve the eigenvalues problem for operator $(-\Delta \hat{+}\alpha (%
\mathbb{-}\Delta _{1})\delta ($\b{x}$))$ we will apply the results of
section 2. First, let us note that $L^{2}(\Omega )$ possess the following
orthogonal decomposition 
\begin{equation*}
L^{2}(\Omega )=\overset{\infty }{\underset{k=1}{\oplus }}\mathcal{H}^{k},%
\text{ \ where }\mathcal{H}^{k}=L^{2}(\mathbb{R}^{2})\otimes \sin kx_{1}.%
\text{\ \ }
\end{equation*}
Further for all $k\in \mathbb{N}$ we have 
\begin{equation*}
\mathcal{H}^{k}=\overset{\infty }{\underset{l=-\infty }{\oplus }}%
L^{2}([0,\infty ),rdr)\otimes e^{ik\phi }\otimes \sin kx_{1}.
\end{equation*}
Then any function $f\in L^{2}(\Omega )$ can be written in the following form 
\begin{equation*}
f(x)=\sum_{k=1}^{\infty }\sum_{l=-\infty }^{\infty }\int_{0}^{\infty }pdp%
\tilde{f}_{lk}(p)J_{l}(pr)e^{ik\phi }\sin kx_{1}
\end{equation*}
where $J_{l}(pr)$ are cylindrical Bessel's functions. Define unitary
operator 
\begin{equation*}
U:L^{2}(\Omega )\rightarrow L^{2}([k^{2},\infty ),d\mu _{lk}),\text{ \ }d\mu
_{lk}(t)=1/2tdt
\end{equation*}
by 
\begin{equation*}
(Uf)_{lk}=\tilde{f}_{lk}(\sqrt{t-\lambda }).
\end{equation*}
The spectral representation of $-\Delta $ has the form 
\begin{equation*}
(U(-\Delta )U^{-1}\psi )_{lk}(t)=t\psi _{lk}(t),\text{ \ \ \ }\psi \in 
\overset{\infty }{\underset{k=1}{\oplus }}\overset{\infty }{\underset{k=1}{%
\oplus }}L^{2}([k^{2},\infty ),d\mu _{lk}).
\end{equation*}
For all $k\in \mathbb{N}$, put $-\Delta ^{k}\equiv -\Delta \lfloor D(A)\cap 
\mathcal{H}^{k}.$ Then $\sigma (-\Delta ^{k})=[k^{2},\infty ).$

Let us note that operator $-\Delta \hat{+}\alpha \mathbb{V}$ satisfies $%
\mathbf{\hat{\sigma}}$ (\ref{ass}) iff for each $k\in \mathbb{N}$ we have 
\begin{equation}
e_{k}\notin Ran(-\Delta ^{k}-E)\text{ \ \ for all }E\in \lbrack k^{2},\infty
).  \label{assexa}
\end{equation}
Proceeding analogously as in the proof of theorem 12 one can show that (\ref
{assexa}) is fulfilled. Next, observing that $e_{k}\in \mathcal{H}^{k}$ for
all $k\in \mathbb{N}$ we get that $\tilde{B}^{-1}$ has the form of (\ref
{invari}). This allows to use theorem 7.

Let us mention that this model corresponds to case 1 described at the end of
section 2. However instead of showing this fact we solve this example
explicitely.

Given $k\in \mathbb{N}$ assume $E_{k}\in (-\infty ,k^{2})$ and define $%
U_{E_{k}0}\equiv U_{E_{k}0}^{k}=-\Delta ^{k}(-\Delta ^{k}-E_{k})^{-1}:$ $%
\mathcal{H}^{k}\rightarrow \mathcal{H}^{k}.$ As follows from theorem 7
number $E_{k}$ belongs to $\sigma _{p}(-\Delta \hat{+}\alpha (\mathbb{-}%
\Delta _{1})\delta ($\b{x}$))$ iff condition\ 
\begin{mathletters}
\begin{equation}
b_{k}^{-1}(U_{E_{k}0}e_{k},e_{k})=E_{k}^{-1}  \label{tuztuz}
\end{equation}
\ holds. A direct calculation yields

\end{mathletters}
\begin{equation*}
(U_{E_{k}0}e_{k},e_{k})=\frac{1}{8\pi ^{4}}\frac{1}{E_{k}}\ln \frac{\alpha
k^{2}}{\alpha k^{2}-E_{k}}.
\end{equation*}
Then, from (\ref{tuztuz}) we get $E_{k}=k^{2}(1-e^{-b_{k}^{\prime }})$ where 
$b_{k}^{\prime }=\frac{1}{8\pi ^{4}}b_{k}.$ By theorem 7 we obtain the
following result.

\begin{corollary}
The discrete spectrum of $-\Delta \hat{+}\alpha (\mathbb{-}\Delta
_{1})\delta ($\b{x}$)$ is given by 
\begin{equation}
E_{k}=k^{2}(1-e^{-2b_{k}})  \label{eigenv}
\end{equation}
where b$_{k}=-\frac{1}{8\pi ^{4}}(\alpha k^{2})^{-1}(1+s_{k}\alpha k^{2}),$\
s$_{k}=\pi ^{-2}(-\ln k/2$+C). The corresponding eigenvectors have the form 
\begin{eqnarray*}
f_{k}=((-\Delta \hat{+}(\mathbb{-}\Delta _{1})\delta (\text{\b{x}}%
))^{-1}U_{E_{k}0}e_{k}=\frac{1}{2}\pi ^{-3}(\int_{\mathbb{R}^{2}}d\text{\b{p}%
}\frac{e^{i\text{\b{p}\b{x}}}}{(p^{2}+k^{2})(p^{2}+k^{2}-E_{k})}\sin kx_{1}
\\
+\frac{1}{2E_{k}}\ln \frac{k^{2}}{k^{2}+E_{k}}\int_{\mathbb{R}^{2}}d\text{\b{%
p}}\frac{e^{i\text{\b{p}\b{x}}}}{(p^{2}+k^{2})}\sin kx_{1}).
\end{eqnarray*}
\end{corollary}

Let us close this example by a short discussion of (\ref{eigenv}). One can
check that $s_{k}<0$\ iff $k\leq 3.$\ So, to describe positive and negative
pure point spectrum we select two cases.

\QTP{Body Math}
Let $k\leq 3.$\ Then $E_{k}>0$\ iff $\alpha \in (0,-k^{-2}s_{k}^{-1}).$

\QTP{Body Math}
Let $k>3$. Then $E_{k}>0$\ iff $\alpha \in (-\infty ,k^{-2}s_{k}^{-1})\cup
(0,\infty ).$

From (\ref{eigenv}) we obtain the following asymptotic behaviour of $%
E_{k}=E_{k}(\alpha )$ as $\alpha \rightarrow -\infty $%
\begin{equation*}
\lim_{\alpha \rightarrow -\infty }E_{k}(\alpha )=k^{2}(1-e^{C}(\frac{2}{k}%
)^{8\pi ^{2}}).
\end{equation*}
Finally, we observe that 
\begin{equation*}
\lim_{k\rightarrow \infty }E_{k}=\infty .
\end{equation*}
Since $\sigma _{ac}(-\Delta )=\sigma _{ac}((-\Delta \hat{+}\alpha (\mathbb{-}%
\Delta _{1})\delta ($\b{x}$))=[1,\infty )$ (see \ref{abcosp}) we get an
interesting result that $E_{k}\in \sigma _{ac}((-\Delta \hat{+}\alpha (%
\mathbb{-}\Delta _{1})\delta ($\b{x}$))$ for sufficiently large \emph{k}
i.e. 
\begin{equation*}
\sigma _{p}((-\Delta \hat{+}\alpha (\mathbb{-}\Delta _{1})\delta (\text{\b{x}%
}))\cap \sigma _{ac}((-\Delta \hat{+}\alpha (\mathbb{-}\Delta _{1})\delta (%
\text{\b{x}}))\neq \{0\}.
\end{equation*}

\subsection{STRONGLY SINGULAR PERTURBATION OF LAPLACE OPERATOR BY THE
DYNAMICS LIVING ON CIRCLE}

Let $\mathcal{H}$ and \emph{A} be as in section 3 i.e. $\mathcal{H}=L^{2}(%
\mathbb{R}^{3},dx)\equiv L^{2},$ $A=-\Delta +\lambda ,$\ $\lambda >0.$\ Then 
$\mathcal{H}_{q}$\ coincides with the Sobolev spaces $W^{2,q}(\mathbb{R}%
^{3})\equiv W^{2,q}$\ (see (\ref{norlap})).\ 

Let us recall that operator $G=A^{-1}$\ can be presented by the integral
kernel 
\begin{equation*}
G(x)=\frac{1}{4\pi }\frac{\exp (-\sqrt{\lambda }\left| x\right| }{\left|
x\right| }=\frac{1}{(2\pi )^{3}}\int_{\mathbb{R}^{3}}dp\frac{e^{ipx}}{%
p^{2}+\lambda }
\end{equation*}
We also introduced notation $G_{r}$ for operator with kernel $G_{r}(x-y)$
given by 
\begin{equation*}
G_{r}(x)=G(x)-\frac{1}{4\pi }\frac{1}{\left| x\right| }.
\end{equation*}

As was shown the spectral representation of \emph{A} determinates the
following decomposition $A=\overset{\infty }{\underset{l=0}{\oplus }}%
\overset{l}{\underset{k=-l}{\oplus }}A^{lk}$ where \ $\sigma
(A^{lk})=[\lambda ,\infty )$ for each \emph{k,l}. Similarly, as in the
general discussion (\ref{homosp}) we define spaces $\mathcal{H}_{k}^{\prime
} $\ as the completions of $D(A)$ in norms 
\begin{equation*}
\left\| u\right\| _{k}^{\prime }=\left\| (A-\lambda )^{k/2}u\right\| _{0}
\end{equation*}
\emph{i.e.} we have 
\begin{equation*}
\left\| u\right\| _{k}^{\prime }=\left\| (-\Delta )^{k/2}u\right\|
_{0}=(\int_{\mathbb{R}^{3}}dx\left| (-\Delta )^{k/2}u(x)\right| ^{2})^{1/2}.
\end{equation*}

We again put $r,$ $\theta ,$\ $\phi $ for the spherical\ \ coordinates in $%
\mathbb{R}^{3}.$ Given real function $\emph{V}\in C(\mathbb{R})$\ define
self-adjoint operator $V(\frac{\partial ^{2}}{\partial ^{2}\phi })$\ in $%
L^{2}(\left\langle 0,2\pi \right\rangle ,d\phi )$\ where $\frac{\partial ^{2}%
}{\partial ^{2}\phi }$\ is the Laplacian with periodic boundary condition.\
For each $k\in \mathbb{Z}\ $put$\ v_{k}=V(k^{2}).$

In this example we will investigate the eigenvalues problem for operator $%
\tilde{A}\in \mathcal{A}_{s}(A)$ which formally corresponds to 
\begin{equation*}
(-\Delta +\lambda )+V(\frac{\partial ^{2}}{\partial ^{2}\phi })\delta
(r-1)\delta (\cos \theta ).
\end{equation*}
Let us put for abbreviation $N=\{r=1,\theta =\frac{\pi }{2},\phi \in
\left\langle 0,2\pi \right\rangle \}$ and $\delta _{c}(r,\theta )\equiv
\delta (r-1)\delta (\cos \theta )$ where the subscript $c$ suggests that the
support of $\delta _{c}$ coincides with the circle in $\mathbb{R}^{3}.$

Define operator

\begin{equation*}
\mathbb{V}_{\alpha }\equiv -\alpha V(\Delta _{\phi })\delta _{c}:D(\mathbb{V}%
_{\alpha })\rightarrow W^{2,-2}
\end{equation*}
by 
\begin{equation*}
\mathbb{V}_{\alpha }f=\sum_{k\in \mathbb{Z}}\alpha v_{k}c_{k}(f)e^{ik\phi
}\delta _{c}(r,\theta );\text{ \ }D(\mathbb{V}_{\alpha })=\{f\in C(\mathbb{R}%
^{3}):\mathbb{V}_{\alpha }f\in W^{2,-2}\}
\end{equation*}
where\ $c_{k}(f)=\int_{0}^{2\pi }d\phi f(r=1,\theta =\frac{\pi }{2},\phi
)e^{-ik\phi }$ and $\alpha \in \mathbb{R}\backslash \{0\}.$

The following facts $Ran\mathbb{V}_{\alpha }\subset W^{2,-2}\backslash
W^{2,-1}$ and $C_{o}(\mathbb{R}^{3}\backslash N)\subset \ker \mathbb{V}%
_{\alpha }$\ ensure that conditions K) and R) are satisfied.

According to general discussion presented in the previous section we
construct operator 
\begin{equation*}
\tilde{A}_{V}\equiv (-\Delta +\lambda )\hat{+}(-\alpha V(\Delta _{\phi
})\delta _{c}).
\end{equation*}
This construction is described in \cite{WKSK}. Below we give the final
result .

Let us abbreviate $\left( p,\phi \right) =p_{1}\cos \phi +p_{2}\sin \phi
+p_{3}$ and put for each $k\in \mathbb{Z}$%
\begin{equation*}
e_{k}=\mathbb{G}e^{ik(\cdot )}\delta _{c}=\frac{1}{(2\pi )^{3}}\int_{\mathbb{%
R}^{3}}dp\int_{0}^{2\pi }d\phi \frac{e^{i(px-\left( p,\phi \right) )}}{%
p^{2}+\lambda }e^{ik\phi }
\end{equation*}
and 
\begin{equation}
q_{k}\equiv c_{k}(G_{r}e^{ik(\cdot )}).  \label{defiqk}
\end{equation}
One can check that $\{e_{k}\}_{k\in \mathbb{Z}}$ is the orthogonal system.

\begin{theorem}[\protect\cite{WKSK}]
Let us assume that $\alpha v_{k}q_{k}\neq -1$ for all k$\in \mathbb{Z}$.
Then operator $\tilde{A}_{V}\in \mathcal{A}_{ss}(A)$ and its inverse is
given by 
\begin{equation*}
\tilde{A}_{V}^{-1}=G+\tilde{B}^{-1}:D(\tilde{B}^{-1})=\{f\in L^{2}(\Omega ):%
\tilde{B}^{-1}f\in L^{2}(\Omega )\}
\end{equation*}
where 
\begin{equation*}
\tilde{B}^{-1}f=\sum_{k=1}^{\infty }b_{k}^{-1}(f,e_{k})e_{k},\text{ \ }%
b_{k}^{-1}\equiv b_{\alpha ,k}^{-1}=-\alpha v_{k}(1+\alpha v_{k}q_{k})^{-1}.%
\text{\ }
\end{equation*}
\end{theorem}

Since this model is more complicated from the technical point of view we
restrict ourselves to some estimations.

First, we will show that this example represents case 2 described at the end
of section 2. Clearly $\tilde{A}_{V}$ satisfies $\mathbf{\hat{\sigma}}.$
Further, let us note that decomposition (\ref{deL2R3}) can be equivalently
written

\begin{equation*}
L^{2}(\mathbb{R}^{3})=\overset{\infty }{\underset{l=0}{\oplus }}\overset{l}{%
\underset{k=-l}{\oplus }}\mathcal{H}^{lk}=\overset{\infty }{\underset{%
k=-\infty }{\oplus }}\overset{\infty }{\underset{l=\left| k\right| }{\oplus }%
}\mathcal{H}^{lk}
\end{equation*}
where $\mathcal{H}^{lk}=L^{2}((0,\infty ),r^{2}dr)\otimes Y_{lk}(\theta
,\phi )$. A direct computation shows 
\begin{equation*}
e_{k}\in \overset{\infty }{\underset{l=\left| k\right| }{\oplus }}\mathcal{H}%
^{lk}\text{ \ \ \emph{for each} }k\in \mathbb{Z}.
\end{equation*}
Therefore $\tilde{B}^{-1}$ has a form of (\ref{invari}). Now, it sufficies
to check that $\mathbb{A}^{1/2}e_{k}\in \mathcal{H}_{-1}^{\prime }$\ for all 
$k\in \mathbb{Z}$. Indeed, we have 
\begin{eqnarray*}
\left\| \mathbb{A}^{1/2}e_{k}\right\| _{-1}^{\prime 2} &=&\int_{\mathbb{R}%
^{3}}dp\int_{0}^{2\pi }d\phi \int_{0}^{2\pi }d\phi ^{\prime }\frac{%
e^{i\left( p,\phi \right) }e^{i\left( p,\phi ^{\prime }\right) }}{%
p^{2}\left( p^{2}+\lambda \right) }\leq \\
4\pi ^{2}\int_{0}^{\infty }dp\frac{1}{p^{2}+\lambda } &=&2\pi ^{3}\lambda
^{-1/2}.
\end{eqnarray*}
So, we get $E_{k}\in \sigma _{p}\left( \tilde{A}_{V}\right) $ iff 
\begin{equation}
b_{k}\in (-\infty ,\lambda \left\| \mathbb{A}^{1/2}e_{k}\right\|
_{-1}^{\prime 2}).  \label{bk}
\end{equation}
Moreover, we have

\begin{equation*}
N_{+}=\#\{b_{k}:0<b_{k}<\lambda \left\| \mathbb{A}^{1/2}e_{k}\right\|
_{-1}^{\prime 2}\},\ \ \ \ N_{-}=\#\{b_{k}:b_{k}<0\}.
\end{equation*}
Note that (\ref{bk}) is equivalent to \ 
\begin{equation}
-\frac{1+\alpha v_{k}q_{k}}{\alpha v_{k}}\leq \lambda \left\| \mathbb{A}%
^{1/2}e_{k}\right\| _{-1}^{\prime 2}.  \label{eicirc}
\end{equation}
A direct calculation using (\ref{defiqk})we get 
\begin{equation*}
q_{k}=-\lambda \left\| \mathbb{A}^{1/2}e_{k}\right\| _{-1}^{\prime 2}.
\end{equation*}
So (\ref{eicirc})\ is satisfied iff $\alpha v_{k}>0.$\ On the other hand for 
$\alpha v_{k}>0$ we have \emph{b}$_{k}<0.$\ Hence 
\begin{equation*}
N_{+}=0,\ \ \ \ N_{-}=\#\{v_{k}:\alpha v_{k}>0\}.
\end{equation*}
Then in this model we have 
\begin{equation*}
\sigma _{p}(\tilde{A}_{V})\cap \sigma _{ac}(\tilde{A}_{V})=\{0\}.
\end{equation*}
\textbf{Aknowledgement }It is pleasure to thank Prof. W. Karwowski for
stimulating discussions and Prof. V. Koshmanenko for valuable remarks. This
paper was supported by 2345/W/IFT/2000.


\begin{thebibliography}{99}
\bibitem{ABK}  S.Albeverio, J. F. Brasche , V.Koshmanenko, \emph{%
Lippman-Schwinger Equation for Singularly Perturbed Operators,} Methods
Func. Anal. \textbf{154}, 130-173,(1998).

\bibitem{AHH}  S.Albeverio, F.Gesztesy, R.Hoegh-Krohn, H. Holden, \emph{%
Solvable Models in Quantum Mechanics,} Springer-Verlag, New York, (1988).

\bibitem{AKK}  S.Albeverio, W.Karwowski, V.Koshmanenko, \emph{Square Power
of Singularly Perturbed Operators}, Math. Nachr. \textbf{173, } 5-24, (1995).

\bibitem{AKK2}  S.Albeverio, W.Karwowski, V.Koshmanenko, \emph{On Negative
Eigenvalues of Generalized Laplace Operator,} submitted for publication.

\bibitem{AlKo}  S.Albeverio, V.Koshmanenko,\emph{\ On the Problem of the
right Hamiltonian under Singular Form-sum Perturbations,} SFB 237 Preprint,
Nr 375, Institut f\"{u}r Mathematik Ruhr-Universit\"{a}t-Bochum, (1997), to
appear in Rev. Math. Phys.\emph{\ }

\bibitem{AK}  S.Albeverio, P.Kurasov, \emph{Singular Perturbation of
Differential Operators}$\emph{,}$ Lon. Math. Soc. Note Series, 271, (2000).

\bibitem{7.aVK}  Yu.Berezanskij, \emph{The bilinear forms and Hilbet
equipment, Spectral analysis of differential operators, Institute of
Mathematics}, Kiev (1980).

\bibitem{Ber}  Yu.Berezanskij, \emph{Selfadjoint operators in Spaces of
Functions of Infinitely Many Variables} (in Russian), Naukova Dumka, Kiev
(1978); English translation: Americam Mathematical Society, Providence
(1986).

\bibitem{BePe}  H. Bethe, R. Peierls,\emph{\ Quantum Theory of Diplon,}
Proc. Soc. London, 148 A, 146-156, (1935).

\bibitem{TKVK}  T.Karataeva, V.Koshmanenko, \emph{Generalized Sum of
Operators, Mathematical Notes}, vol. \textbf{66},No5, (1999).

\bibitem{WKSK}  W.Karwowski, S.Kondej, \emph{The Laplace Operator, Null Set
Perturbation and Boundary Conditions, } submitted for publication.

\bibitem{KKO}  W.Karwowski, V.Koshmanenko, S.\^{O}ta, \emph{Schr\"{o}dinger
operator perturbed by operators related to null-sets}, Positivity \textbf{2}%
, no. 1 (1998), 77-99.

\bibitem{SK1}  S.Kondej, \emph{Singular Pertrubation of Laplace Operator in
the terms of Boundary Conditions}, submitted for publication.

\bibitem{VK3}  V.Koshmanenko, \emph{Singular Bilinear Forms and Self-adjoint
Extensions of Symmetric Operators}, Spectral Analysis of Differential
Operators (in Russian), Institute of Mathematics, Kiev, 37-48 (1980).

\bibitem{VK2}  V.Koshmanenko, \emph{Perturbation of self-adjoint operators
by singular bilinear forms}, Ukrainian Math. J.\textbf{43}, no.11 (1991),
1559-1566.

\bibitem{VK1}  V.Koshmanenko, \emph{Singular Operator as a Parameter of
Self-adjoint Extensions}, Op.Th. vol. \textbf{118}, (2000).

\bibitem{KrYa}  M. G. Krein, V. A. Yavryan,\emph{\ Spectral Shift Functions
that arise in Perturbations of a Positive Operator,} J. Operator Theory 
\textbf{6}, 155-191 (1981).

\bibitem{KrPe}  R. de L. Kr\"{o}nig, W.G. Penney, \emph{Quantum Mechanics of
Electron in Crystal Lattices,} Proc. Soc. London, vol \textbf{130,} 499-513,
(1931).

\bibitem{Ma}  V.G. Maz'ya, \emph{Sobolev Spaces,} Springer, Berlin New\
York, 1985.
\end{thebibliography}
\end{document}